\documentclass[useAMS,usenatbib]{mn2e}
\usepackage{latexsym}
\usepackage{makeidx}
\usepackage{cite}
\usepackage{amsmath}
\usepackage{color}
\usepackage{upgreek}
\usepackage{morefloats}
\usepackage{lineno}
\usepackage{setspace}
\usepackage{graphicx}

\title[Telling twins apart: Exo-Earths and Venuses with transit spectroscopy]{Telling twins apart: Exo-Earths and Venuses with transit spectroscopy}
\author[J.~K. Barstow et al.]{J.~K. Barstow$^{1,2}$\thanks{E-mail:
jo.barstow@astro.ox.ac.uk (JKB)}, S. Aigrain$^{1}$,
P.~G.~J. Irwin$^{2}$, S. Kendrew$^{1}$, L.~N. Fletcher$^{3}$\\
$^{1}$Astrophysics, Denys Wilkinson Building, Department of Physics, University of Oxford, UK\\
$^{2}$Atmospheric, Oceanic and Planetary Physics, Clarendon Laboratory, Department of Physics, University of Oxford, UK\\
$^{3}$Department of Physics and Astronomy, University of Leicester, UK}
\begin{document}

\date{Submitted *** 2015}

\pagerange{\pageref{firstpage}--\pageref{lastpage}} \pubyear{2015}

\maketitle

\label{firstpage}

\begin{abstract}
The planned launch of the \textit{James Webb Space Telescope} in 2018 will herald a new era of exoplanet spectroscopy. \textit{JWST} will be the first telescope sensitive enough to potentially characterize terrestrial planets from their transmission spectra. In this work, we explore the possibility that terrestrial planets with Venus-type and Earth-type atmospheres could be distinguished from each other using spectra obtained by \textit{JWST}. If we find a terrestrial planet close to the liquid water habitable zone of an M5 star within a distance of 10 parsecs, it would be possible to detect atmospheric ozone if present in large enough quantities, which would enable an oxygen-rich atmosphere to be identified. However, the cloudiness of a Venus-type atmosphere would inhibit our ability to draw firm conclusions about the atmospheric composition, making any result ambiguous. Observing small, temperate planets with \textit{JWST} requires significant investment of resources, with single targets requiring of order 100 transits to achieve sufficient signal to noise. The possibility of detecting a crucial feature such as the ozone signature would need to be carefully weighed against the likelihood of clouds obscuring gas absorption in the spectrum. 
\end{abstract}

\begin{keywords}
Methods: data analysis -- planets and satellites: atmospheres -- radiative transfer
\end{keywords}

\maketitle

\section{Introduction}
The field of exoplanet science has been transformed over the last few years, with discoveries of water vapour (e.g. \citealt{wakeford13,madhu14}), alkali metals (e.g. \citealt{redfield08}) and even clouds (e.g. \citealt{pont13,kreidberg14}) in the atmospheres of exotic worlds. The majority of these discoveries were made using the transit spectroscopy technique. When a planet transits in front of its parent star, it blocks a small fraction of the starlight; gases or particulates in planet's atmosphere absorb or scatter starlight at particular wavelengths, meaning that the fraction of light blocked varies as a function of wavelength. 

Currently, the majority of transiting exoplanets for which we have atmospheric information are hot Jupiters, because their size and extended atmospheres produce a large transit spectroscopy signal. Reported discoveries of `Earth-like planets' from the \textit{Kepler} mission rest on assumptions based on their measured sizes, and estimates of the planets' temperatures from their orbital periods. However, the conditions on a planet's surface are dependent on the atmospheric properties as well as the distance from the star, and in particular calculations of the extent of the solar system liquid water habitable zone (HZ) are model-dependent and show some disagreement. Recent calculations by \citet{kopparapu13} using a 1D, cloud-free climate model indicate that the Earth is very close to the inner edge of the HZ, which in this model is located 0.99 AU from the Sun, and Venus is well outside it; however, modelling by \citet{zsom13} indicates that for certain conditions of humidity, surface pressure and surface albedo the HZ could extend as far in as 0.38 AU. In addition, 3D models presented by \citet{leconte13} indicate that tidally-locked planets that might otherwise experience a runaway greenhouse may actually have stable liquid water in cold traps on the nightside.

Estimating habitability based on orbital period and planet size might work in our own solar system today, but Venus is approximately Earth-sized and may have been in the HZ as recently as 1 billion years ago \citep{abe11}. Therefore, it involves risky assumptions for exoplanetary systems. However, the development of larger and more sensitive telescopes over the next few years should make it easier to target smaller, denser planets for spectroscopic follow-up observations. Whilst the spectral signals of interest for Earth-size planets transiting sun-like stars are too small relative to the star's brightness, of order 1:1,000,000, Earth-sized planets around smaller M dwarf stars may have large enough signals. In addition, the much smaller orbital radius of the habitable zone around these cool stars means temperate planets have short periods, making it easier to combine multiple observations to boost the signal. 

The \textit{James Webb Space Telescope (JWST)}, due to launch in 2018 with a 25 m$^2$ primary mirror, may be capable of characterising the atmosphere of an Earth-sized planet orbiting an M dwarf, should such a system be found in our near neighbourhood (within 10 pc). Surveys such as TESS \citep{ricker15} and the Next Generation Transit Survey \citep{wheatley13}, which are optimised for detecting transits around red stars, may discover such planets. \citet{deming09,kaltenegger09,barstow15} showed that it might be possible to detect ozone in the atmosphere of an Earth analog planet in orbit around an M dwarf, but full characterisation of the atmosphere's composition and structure {is likely to be difficult even for a favourable case. However, detection of ozone is in itself of interest, especially for a planet in the habitable zone of its parent star, as it might indicate that the planet has an Earth-like atmosphere. Simulating Venus in transit (as e.g. \citealt{ehrenreich12}) indicates which features might be observable for this more inhospitable world. 

In this paper, we investigate whether it would be possible to distinguish between an Earth-analog planet and a Venus-analog planet orbiting an M dwarf, using transmission spectra obtained by \textit{JWST}. At first glance, the atmospheres of Earth and Venus are very different, but in transmission geometry light penetrates to a limited depth in the atmosphere, especially if clouds are present (e.g. the case of super-Earth GJ 1214b, \citealt{kreidberg14}). Both planets are cloudy, so this may mask their atmospheric composition differences in transmission spectra. Secondary transit spectra for these planets have low signal-to-noise, even when averaged over several tens of eclipses (Figure~\ref{secondaries}).

\begin{figure*}
\centering
\includegraphics[width=0.85\textwidth]{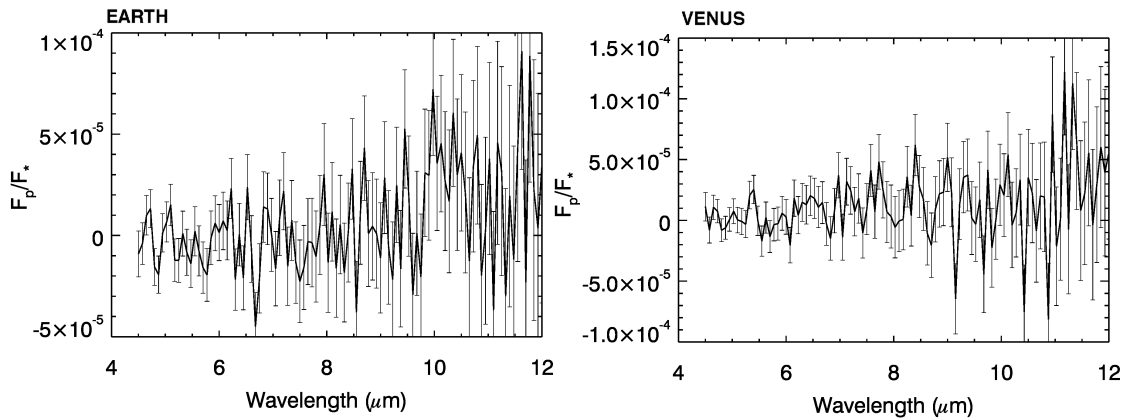}
\caption{Secondary eclipse spectra with JWST/MIRI for Earth and Venus around an M5 star at 10 pc. Spectra are averaged over 30 (50) eclipses for Earth (Venus). The low signal to noise means that very little information is obtainable from these spectra.\label{secondaries}}
\end{figure*}

\section{Earth and Venus}
Earth and Venus are the fraternal twins of the solar system. They are similarly-sized rocky bodies (Venus has a radius of 0.95R$_{\oplus}$), and both exist either within or close to the liquid water habitable zone. Even though Venus is closer to the Sun, because it is highly reflective it has a slightly lower equilbrium temperature of $\sim$250 K compared with Earth's 260 K. 

However, the similarities end here. Venus has a much higher surface temperature and pressure than Earth, 735 K and 92 bars respectively, compared with around 287 K and 1 bar for Earth. At some point in its past, Venus's atmosphere entered a runaway greenhouse phase, during which the majority of its water was lost \citep{rasool80,donahue82}. Without any surface water to facilitate carbonate rock formation, almost all of Venus's carbon budget is in the atmosphere in the form of CO$_2$, making up 96 \% of the total atmospheric volume. This accounts for the extreme surface conditions.  Due to the loss of water Venus's atmosphere is very dry compared with Earth's, but it does have cloud and haze made of sulphuric acid (H$_2$SO$_4$, \citealt{hanhov,pollack74}). The sulphuric acid aerosols are formed as a photochemical product of SO$_2$ gas and the little water vapour that is present. Venus's atmosphere has high concentrations of sulphur gases such as SO$_2$ and OCS, probably as a result of significant past or current volcanic activity. It is however lacking in bio-activity-related gases O$_2$ and ozone, which are present in significant quantities on Earth (see \citealt{taylor09} for a summary of Venusian climate). 

A detailed understanding of their atmospheres is needed to reveal the very different surface conditions on Venus and Earth. Identifying an Earth analog planet using transit spectroscopy will therefore require Venus-like scenarios to be ruled out. It may be possible, using the soon-to-be-launched \textit{James Webb Space Telescope}, to identify absorption features due to key gases and distinguish between Venus-like and Earth-like atmospheres. 

\section{Synthetic spectra}
Synthetic spectra for Earth and Venus were produced using the \textit{NEMESIS} radiative transfer and retrieval code \citep{irwin08}. \textit{NEMESIS} contains a 1D radiative transfer model that incorporates a rapid correlated-k (\citealt{lacis91}, after \citealt{goodyyung}) radiative transfer scheme, and an optimal estimation retrieval algorithm \citep{rodg00}. Properties specified in the forward models include temperature and atmospheric gas abundances as a function of pressure, cloud specific density and wavelength-dependent extinction cross-section, and planet mass and radius. The stellar radius is also specified and becomes important in the calculation of transit depth.

The analog planets are based on the current atmospheric compositions of Earth and Venus; we assume that they evolved in exactly the same way as Earth and Venus did in the solar system, so they have identical atmospheric characteristics. The validity of this assumption, and possible alternatives, are explored in Section~\ref{discussion}. The Earth model used is as presented in \citet{irwin14}, and the Venus model is based on that used in \citet{barstow12}, after \citet{seiff85}. Line data for both planets are taken from the HITRAN08 database \citep{rothman09}. Sulphuric acid refractive indices for the Venus cloud model are taken from \citet{pw} ($\lambda <$ 5 $\upmu$m) and \citet{myhre} ($\lambda >$ 5 $\upmu$m). Whilst it is usually necessary to use high temperature line databases for Venus, due to the long path length in transmission geometry transit spectra are insensitive to the atmosphere below the 1 bar level, which on Venus corresponds to around 50 km. The high temperature portion of the Venusian atmosphere is effectively inaccessible and therefore high temperature line data are not required in this case. 

Both models are based on the best-fitting atmospheric composition, structure and cloud properties for each planet. The only exception is the introduction of a reduced cloud opacity model for Venus, to test the sensitivity to the presence of cloud. Whereas clouds on Earth are patchy and the cloud top pressure is rarely lower than 100 mbar (e.g. measurements from the SEVIRI satellite analysed by \citealt{hamann14}), the cloud layer on Venus is permanently present and extends up to the 1 mbar level (see \citealt{esposito83} for a summary of the major features). This means that the higher, thicker Venusian cloud will have a much larger impact on transmission spectra than the Earth cloud. An example of the effect of increasing cloud top altitude on transmission spectra is shown by \citet{betremieux14}, with increasingly higher cloud top altitudes for an Earth-like planet resulting in smaller and smaller molecular absorption features; \citet{barstow13b} shows the same effect for the super-Earth exoplanet GJ 1214b.

The planets are assumed to be in orbit around a 3000 K M5 red dwarf with a radius of 0.14R$_{\odot}$ and a mass of 0.122M$_{\odot}$. We calculate that, for the planets to have the same equilibrium temperature as they do in the Solar System, Venus and Earth would be in orbits of $\sim$5 and 8 days respectively. This would make it possible to observe the planets in transit many times during the lifetime of \textit{JWST}. 

\section{Noise model}

Noise is added to the spectra using the same methods and instrument characteristics as those presented in \citet{barstow15}. As in the previous work, the simulated spectra are based on a prism mode observation with the NIRSpec instrument, covering 0.6 to 5 $\upmu$m, and a MIRI Low Resolution Spectograph observation covering 4.5 to 12 $\upmu$m. In this paper, we do not consider the effect of systematic errors and starspots; we refer the reader to \citet{barstow15} for a detailed discussion of these issues for \textit{JWST}, and discuss the possible impacts on this case in Section~\ref{discussion}.

The noise for both instruments is calculated assuming the photon noise limit is reached, using the equation

\begin{equation}
n_{\lambda}=I_{\lambda}{\times}{\pi}{\times}(r_{\star}/D_{\star})^2{\times}({\lambda}/hc){\times}({\lambda}/R){\times}A_{eff}{\times}QE{\times}{\eta}{\times}t\label{eqn1}
\end{equation}

where $n_{\lambda}$ is the number of photons received for a given wavelength $\lambda$, $I_{\lambda}$ is the spectral radiance of the stellar signal, $r_{\star}$ is the stellar radius, $D_{\star}$ is the distance to the star, $h$ and $c$ are the Planck constant and speed of light, $R$ is the spectral resolving power, $A_{eff}$ is the telescope effective area, $QE$ is the detector quantum efficiency, $\eta$ is the the throughput and $t$ is the exposure time. For NIRSpec, we adopt the average transmission and quantum efficiency properties used by \citet{deming09}, assuming a detector QE of 0.8, a telescope optics efficiency of 0.88 and a total NIRSpec optics transmission of 0.4. For MIRI-LRS, we use the photon conversion efficiency presented in \citet{kendrew14}. This is the fraction of photons from the source that are eventually received and recorded by the detector, combining the QE and throughput. The total area of the primary mirror is 25 m$^2$.  Exposure time is equal to the transit duration across the stellar equator for the assumed orbits, with an assumption of an 80 \% duty cycle as used in previous work \citep{barstow13,barstow15}.

Although the planet:star size ratio is favourable for these cases, M5 stars are cool and therefore relatively faint. Even for a nearby star, multiple observations would be required with \textit{JWST}. The models presented assume 30 observations (50 for Venus) each with MIRI and NIRSpec, for planets orbiting an M5 star at 10 pc distance. MIRI and NIRSpec cannot be used to observe simultaneously, so this translates to 60 (100) observations per planet in total, of around 3 hours each including the out-of-transit baseline. This would be easily feasible in the lifetime of JWST due to the short orbital periods of both planets, but it would clearly be a major investment of valuable observational time.

These values can be compared with the condition adopted by \citet{deming09} that 60 transits  would be reasonable to characterise a habitable 10M$_{\oplus}$ super Earth. The size of observable atmospheric signals remains relatively constant from Earth through to super Earth-sized objects, since the effects of increased gravity/smaller atmospheric scale height offset the increased area of the transiting planet. \citet{deming09} estimate that TESS would discover between 1 and 5 habitable super-Earths for which H$_2$O and CO$_2$ absorption could be characterised using JWST. \citet{kaltenegger09} found that multiple transits would be required to detect O$_3$ in the atmosphere of an Earth-analog planet around an M dwarf, finding a SNR of 20 could be achieved for the infrared O$_3$ feature with 200 hours of in-transit observation. 

\section{Retrieval tests}

The aim of this study is to test whether it is possible to distinguish between Earth-like and Venus-like atmospheres of terrestrial exoplanets. In the majority of cases, where the planet is at least partially cloudy, it is virtually impossible to recover the full atmospheric state from a transit spectrum due to degneracies between bulk composition, cloud and temperature (see e.g. \citealt{benneke12,barstow13, barstow13b}, and \citealt{griffith14} for a useful summary). Therefore, what we are interested in is not an accurate recovery of the atmospheric state vector, rather it is whether or not the synthetic observation can best be matched with a Venus- or Earth-like model atmosphere. It is of course also possible that an Earth-sized object in another solar system may have evolved an atmosphere different from either of these; however, for purposes of comparative planetology, a key goal following the discovery of another terrestrial habitable-zone planet would be to see whether it most closely resembles the temperate, habitable scenario found on Earth, or the Venus runaway greenhouse.  

To this end, we perform retrieval tests for each planet with a range of nine model atmospheres, ranging from Venus-like to Earth-like with intermediate cases in between. Since there is little sensitivity to temperature in primary transit spectra, we use the same temperature profile (Figure~\ref{tempprof}) in the retrieval model for both cases. The temperature mainly affects the scale height, which impacts the size but not the presence of gas absorption features, and Venus and Earth actually have broadly similar temperature-pressure profiles where they overlap in pressure, although the structure differs. This makes the test somewhat more realistic, as information about the temperature structure can only be obtained from secondary transit observations. Otherwise, it must be estimated ab initio from the equilibrium temperature, which would result in a similar profile to the one used. As shown in Figure~\ref{secondaries}, secondary transit signal-to-noise for these objects is too small to obtain any useful information about temperature structure.

\begin{figure}
\centering
\includegraphics[width=0.5\textwidth]{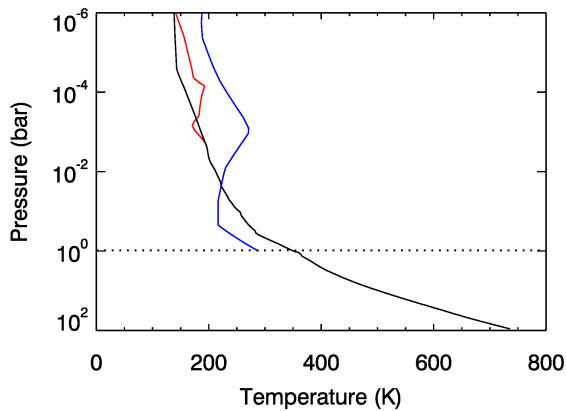}
\caption{The temperature-pressure profile used in the retrievals (black) compared with the input profiles for Earth (blue) and Venus (red). The surface for the Earth profile is indicated by the dashed line. The profile used for the retrievals is a smoothed version of the Venus profile. \label{tempprof}}
\end{figure}

We include a combination of the gases expected to be found in Earth and Venus atmospheres in the retrieval models; H$_2$O, CO$_2$, O$_3$, CO, CH$_4$, O$_2$, SO$_2$, OCS, and N$_2$. We test a set of nine retrieval model atmospheres, ranging from the Earth case (78\% N$_2$, 21\% O$_2$, negligible SO$_2$ and OCS, significant O$_3$) to the Venus case (96\% CO$_2$, 3.5 \% N$_2$, negligible O$_2$ and O$_3$, including SO$_2$ and OCS). Seven intermediate models have intermediate abundance of each gas with even steps in log-space, except for N$_2$ which has no absorption lines and is used to bring the abundance total up to 100\%. The abundance priors for all gases except N$_2$, for each of the nine models, are shown in Figure~\ref{gases}. Because several gases are not included at all in the input model for either Earth or Venus, the \textit{a priori} abundances for the retrieval can be as low as 10$^{-36}$. For these cases, where a gas is only considered to be present in either the Earth or Venus like atmosphere, the gas only has an observable effect on the spectrum close to either the Earth or Venus model end members. The synthetic spectra correspoding to each of the nine models are shown in Figure~\ref{synthetics}.

\begin{figure}
\centering
\includegraphics[width=0.5\textwidth]{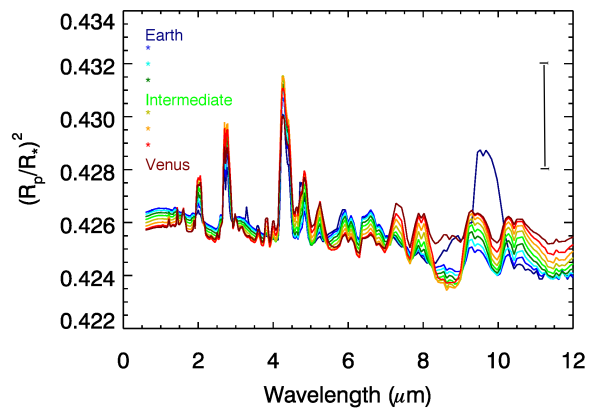}
\caption{Synthetic spectra (for Earth bulk properties) for each of the nine atmospheric composition models. Note that the ozone feature at 9 $\upmu$m is only large enough to be seen in the most Earth-like case (navy blue line), and an SO$_2$ can be seen in the dark red Venus-like spectrum at 8.5 $\upmu$m; however, this second feature is on the order of the noise, so in practice is unlikely to be observable. Noise is indicated by the bar on the right of the plot.\label{synthetics}}
\end{figure}

For each planet, we run a separate retrieval with each of the nine atmospheric priors. We allow cloud optical depth and abundances of H$_2$O, CO$_2$, O$_3$, CO, CH$_4$ and O$_2$ to vary in the retrieval, and we retrieve scaling factors for all these quantities. The cloud prior for Venus is discussed later in Section~\ref{results}.  Precise abundances of SO$_2$ and OCS did not have a significant effect on the spectra, and in fact none of the retrieved gas abundances deviated very far from the \textit{a priori} case during the retrieval, suggesting that there is little sensitivity to the precise abundance of each gas; however, as discussed in Section~\ref{results}, we find that there is sensitivity to the presence or absence of certain gases. 

 We also retrieved the radius at the base of the atmosphere, in these cases at the planet's surface. The reason for this is that the radius at a given pressure is not known for an exoplanet until the atmospheric state is also known; the pressure being probed at the white light transit radius depends on the gas absorbers present in the atmosphere and also the cloud top pressure. The molecular weight of the atmosphere depends on the model atmosphere composition and is calculated within \textit{NEMESIS}. 

For the Earth retrievals, we use an Earth-like water cloud model as included by \citet{irwin14}, which has relatively little effect on the retrieval due to the low optical depth and deep cloud top. We test two different cloud models for the Venus retrieval - a Venus-like model based on \citet{barstow12} with optically thick cloud up to $\sim$80 km altitude, and a reduced optical depth version of the same model which would have a much smaller effect on the spectrum.The reduced-OD model has the upper cloud optical depth reduced by a factor of 10,000 from cloudy Venus model, representing a significant clearing of the cloud.

\begin{figure*}
\centering
\includegraphics[width=0.85\textwidth]{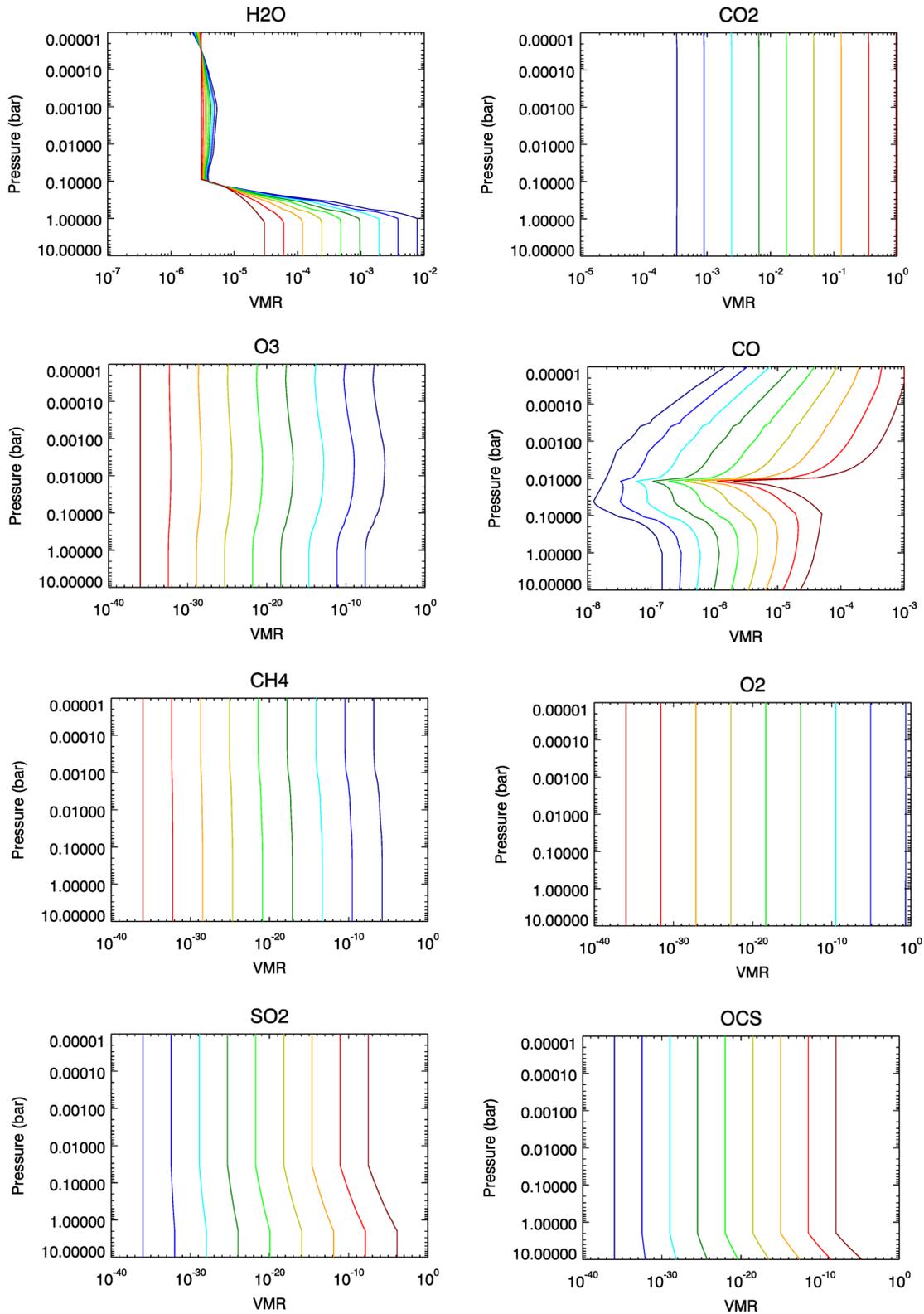}
\caption{Gas abundance profiles for the Earth/Venus model atmospheres. The abundance of each gas is expressed as a volume mixing ratio (VMR), defined as the molecular number density of each gas over the total number density of atmospheric molecules. The dark blue case is the Earth-like atmosphere, dark red is the Venus-like atmosphere, and the intervening models follow the rainbow sequence with lime green marking the 50:50 case. The abundance of H$_2$O in the stratosphere does not vary significantly between the two planets due to the condensation to form water clouds on Earth mimicking the dryness of the Venusian atmosphere.\label{gases}}
\end{figure*}

\section{Results}
\label{results}
We present results from the Earth test and from three Venus tests with different cloud properties and priors. $\chi^2$ values, the synthetic observation and fitted spectra for Earth are shown in Figure~\ref{earthspec}, and for Venus in Figure~\ref{venusspec}. The number of degrees of freedom (measurements - number of variables) is 251. 

\begin{figure*}
\centering
\includegraphics[width=0.85\textwidth]{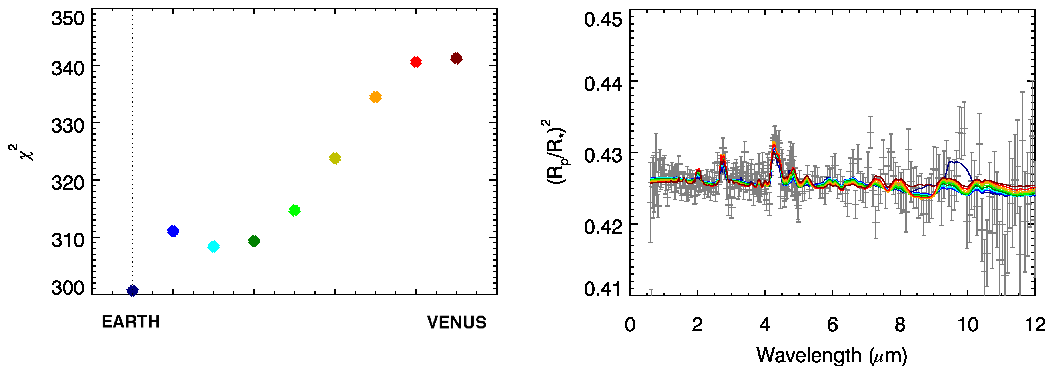}
\caption{$\chi^2$ goodness of fit measurements and fitted spectra for the Earth case. The lowest $\chi^2$ is marked by a dashed line. The Earth model (dark blue), containing a significant amount of ozone, clearly provides the best fit to the synthetic observation. The location of the ozone feature in the spectrum is indicated by a dashed line.\label{earthspec}}
\end{figure*}

\begin{figure*}
\centering
\includegraphics[width=0.85\textwidth]{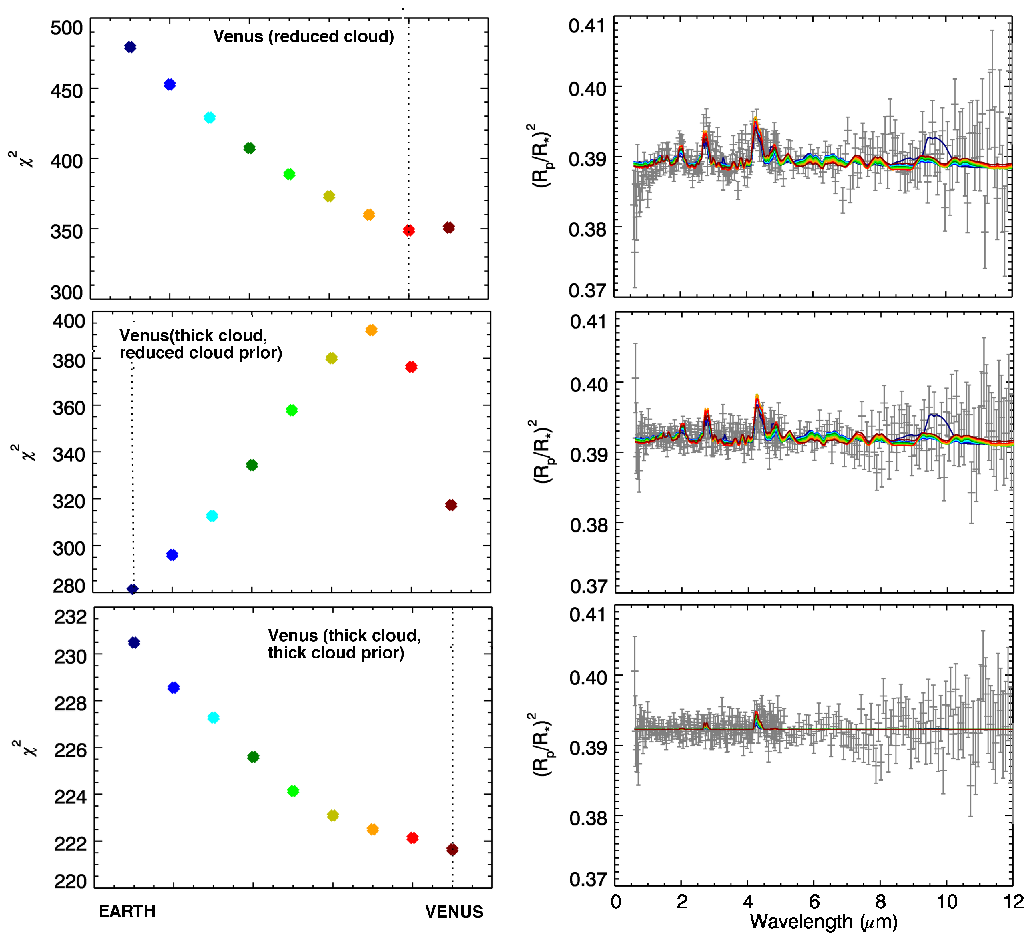}
\caption{$\chi^2$ goodness of fit measurements and fitted spectra for the reduced-cloud Venus model (top), the cloudy Venus model with a reduced-cloud prior (middle) and the cloudy Venus model with a cloudy prior (bottom). The best-fitting model in each case is indicated by a dashed line on the $\chi^2$ plot. For the cases where the cloud prior used matched the true conditions well, a better fit is found for Venus-type model atmospheres, but in the case where the prior underestimates the cloud optical depth (middle plots) the spectrum would be incorrectly identified as Earth-like. \label{venusspec}}
\end{figure*}

In the Earth case, the Earth-like, ozone-containing model provides a much better fit than any other, with the lowest $\chi^2$ (Figure~\ref{earthspec}). This is because the ozone feature at 9 $\upmu$m is clearly seen in the synthetic observation, and cannot be reproduced by combinations of other gases. This supports the findings briefly presented in \citet{barstow15}, that the ozone feature could be detected in the atmosphere of an Earth twin around a small, cool star.

However, the case is less clear-cut for Venus. For the cloudy Venus-twin case with the cloudy model prior, the Venus model does produce the best fit, but the range of $\chi^2$ values is far narrower than for the Earth case. This is because the thick cloud, extending up to the low pressure region of the atmosphere, cuts off the majority of the gas absorption features; only the centre of the 4.3 $\upmu$m CO$_2$ band is visible above the noise. This means the information available from the spectrum is very limited, so little constraint is provided on the properties of the atmosphere.

Due to the extreme flatness of the spectrum, if a reduced-cloud \textit{a priori} model is used to fit the cloudy Venus spectrum, something quite strange happens (middle row, Figure~\ref{venusspec}). The model with the lowest $\chi^2$ is actually the Earth atmosphere model. Despite the fact that this model contains an ozone feature where none is seen in the spectrum, it fits well because the size of the 4.3 $\upmu$m CO$_2$ absorption feature matches most closely. The CO$_2$ feature in the synthetic observation is small, because the cloud prevents the deep wings of the feature from being seen. However, in the Earth model, the CO$_2$ feature is also relatively small because the CO$_2$ abundance is smaller than in any of the other models, hence the Earth model provides the best fit (Figure~\ref{venusspec}, middle row).

For the reduced-cloud Venus synthetic observation, Venus-like models provide the best fit to the spectrum, but there is still more ambiguity than there is for the Earth case as the best-fitting model has some Earth-like characteristics. The reason for this is the lack of an ozone-like feature - there is no gas that is present on Venus and not on Earth that also has an unmistakeable absorption feature. Whilst SO$_2$ and OCS fulfil the first part of this criterion, the absorption features are not strong enough to be identified above the noise in the transmission spectra, unlike the Earth O$_3$ feature. This suggests that, whilst it might be possible to unambiguously spot an ozone-rich planet, detecting an obvious Venus is harder. 

We show the size of significant gas features relative to the noise level for the Earth and reduced-cloud Venus cases in Figure~\ref{features}, which demonstrates that the Earth case has stronger unique features; the only very strong features for Venus are CO$_2$ absorption bands, which are also found in Earth's atmosphere. 250 transits of a Venus-like planet would be required to detect the 6---7 $\upmu$m H$_2$O feature, the largest feature after the CO$_2$ bands, and even then SO$_2$, a key Venusian gas, is still undetectable. This is partly due to the effect of clouds, but also the fact that the main atmospheric constituent is CO$_2$, which has a high mean molecular weight and many absorption features that mask the signatures of other gases. The situation is of course even worse for the cloudy-Venus scenario, and clouds are likely to play a crucial role in the feasibility of constraining Earth-like atmospheres.

\section{Discussion}
\label{discussion}
The results presented above rely on the validity of several assumptions made during the modelling process. We consider these assumptions here, and discuss their implications for real observations of M dwarf terrestrial planets. 

\subsection{Photochemistry and the effect of the M dwarf primary}
Throughout this paper, we have assumed that it is possible for two rocky planets close to the liquid water habitable zone of an M5 star to evolve exactly as Venus and Earth did in the solar system. However, since it is known that photochemistry has played an important role in the evolution of both atmospheres, and M dwarfs have very different UV fluxes and spectra to G-type stars, this assumption may be flawed. In particular, the formation of the ozone layer on Earth, which is the atmosphere's most distinctive feature in infrared transmission, is entirely dependent on UV photochemical processes. How likely is it that a similar ozone layer would have formed on an oxygen-rich planet orbiting a cooler star?

\begin{figure}
\centering
\includegraphics[width=0.5\textwidth]{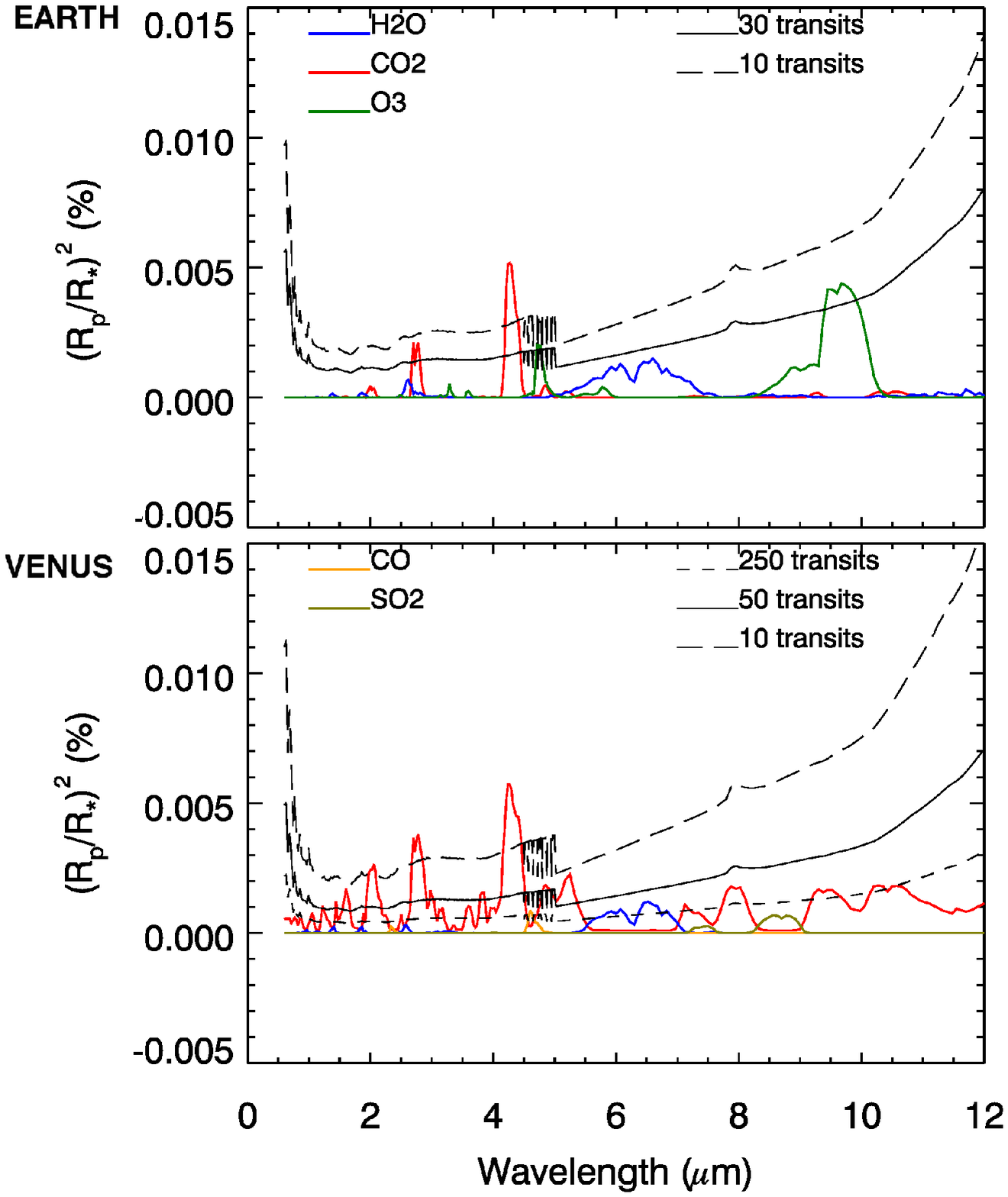}
\caption{The size of gas absorption features relative to the noise level for the Earth (top) and reduced-cloud Venus (bottom) cases. The sizes of the features are calculated by removing each gas from the full spectrum one at a time and taking the difference between the spectra. For both planets, features are small at short wavelengths as these wavelengths are dominated by clouds and scattering to a greater extent. All features other than CO$_2$ appear weak for Venus. Around 30 transits is the minimum required to detect O$_3$ for the Earth case.\label{features}}
\end{figure}

\citet{segura05} showed that whilst O$_3$ production would still be expected around a cooler star, the chemical production and destruction processes are very different and the balance is not the same. \citet{grenfell13} suggest that the O$_3$ column abundance around a  3100 K star would be decreased by more than a factor of 3 from the expected abundance around a sun-like star; this is mainly due to the Chapman scheme, the photochemical process responsible for producing the majority of Earth's ozone, becoming much less efficient. Instead, production due to smog processes dominates, but ozone loss due to the presence of NOx gases and oxidation of CO is likely to become more efficient, resulting in a net decrease in the O$_3$ column abundance. \citet{segura05} find that the UV flux profile of an active AD Leo-type star would result the O$_3$ column abundance being reduced by around 50\%, but for a star like GJ 643C it could be increased by the same amount. \citet{rugheimer15} find that for models of active M dwarf stars the O$_3$ column abundance remains relatively high even for later stellar types, due to their high UV fluxes, whereas for an inactive model M5 star the column abundance is decreased by a factor of 20 from the solar value of $\sim$9$\times$10$^{18}$ cm$^{-2}$ used in our model. Generally, for quiescent stars, the O$_3$ column depth decreases gradually towards later spectral types. However, the inactive stellar models are presented by \citet{rugheimer15} as a limiting case; of the stars with measured UV fluxes discussed in the paper,  GJ 876 has a temperature closest to an M5 spectral type star, and \citet{rugheimer15} find that a planet orbiting this star would have an O$_3$ column abundance of 8.8$\times$10$^{17}$ cm$^{-2}$, nearly double that of the inactive M5 model.

In Figure~\ref{earthspec2}, we demonstrate the effect of a reduced ozone column abundance on our ability to distinguish an Earth-like atmosphere from a Venus-like atmosphere. We run the retrieval test with a synthetic observation based on a model with 0.3$\times$ the Earth O$_3$ abundance, as indicated by \citet{grenfell13}. It is clear that while the ozone-rich Earth model still provides the best fit to this synthetic observation, it is less straightforward to distinguish the model containing substantial O$_3$ from other Earth-like models. Assuming the same concentration of oxygen as in Earth’s atmosphere, the lower UV environment around M stars will lead to decreased photochemical production of ozone (e.g. \citealt{segura05,rauer11,grenfell13,rugheimer15}). Therefore detecting some biosignatures in transit, such as ozone, may be more difficult \citep{deming09,kaltenegger09}, while others, such as methane, may build up to more detectable levels in the atmosphere \citep{segura05,rugheimer15}. Some of this effect may perhaps be compensated for by increasing the number of transits observed, but this may introduce other problems (see Section~\ref{temporal_changes}).

However, photochemistry may not be all bad news for detecting biosignatures around M dwarf planets. In lower UV environments certain biogenic gases such as CH$_4$, N$_2$O and CH$_3$Cl have much longer lifetimes and therefore may have higher abundances on those planets for the same biogenic fluxes \citep{segura05,rauer11,grenfell13,rugheimer15}, which may mean they would be observable. CH$_4$ in particular could have a mixing ratio of up to 200 times the Earth value, which produces significant changes in the transmission spectrum (Figure~\ref{ch4inc}). A combination of a reducing biogenic gas and high oxygen abundance indicated by O$_3$ would be a more convincing indication of the presence of life than a detection of O$_3$ alone.

\begin{figure*}
\centering
\includegraphics[width=0.85\textwidth]{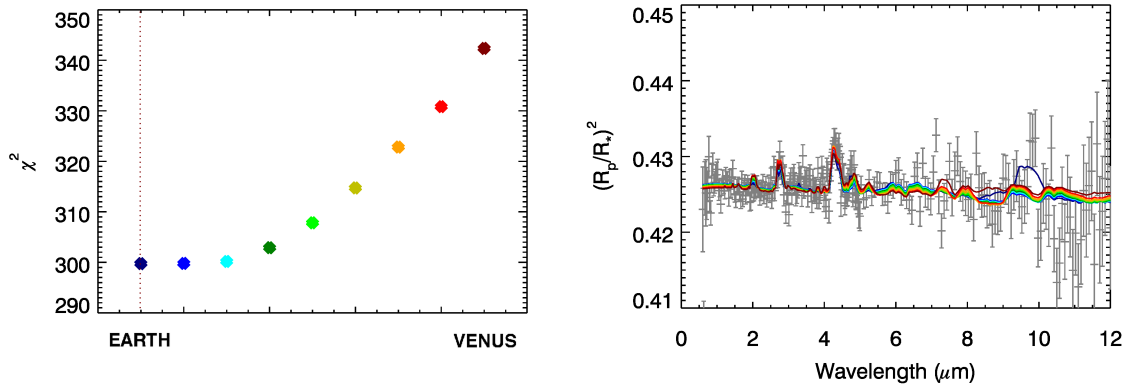}
\caption{$\chi^2$ goodness of fit measurements and fitted spectra for the Earth case, where the ozone abundance in the synthetic observation is reduced to 0.3$\times$ the true Earth abundance. The Earth model (dark blue) still provides the best fit, but only just. More Earth-like models clearly provide a better fit to the spectrum than Venus-like models, but the ozone absorption signal is no longer large enough to distinguish the true-Earth model from other Earth-like models. \label{earthspec2}}
\end{figure*}

Cosmic rays from flaring M dwarf stars can further change the mix of gases present in a planet's atmosphere. \citet{tabataba15} show that cosmic rays can suppress NOx gas destruction of O$_3$, and increase HOx destruction of CH$_4$, indicating that long-term observations of M dwarf planet hosts are required. \citet{segura10} show that including the effect of protons from M dwarf flares on calculations of O$_3$ column abundance has significant effects. The detection of flares in a stellar light curve might have a strong bearing on the interpretation of gas abundances measured from transmission spectra. Of course, flaring M dwarf stars may not be the best candidate hosts for Earth-like planets anyway, as the flares may be problematic for the evolution of life on close-in planets. This means that the finding by \citet{rugheimer15} that planets orbiting inactive M dwarfs are likely to have weaker O$_3$ signatures than those orbiting active ones poses some problems for detecting O$_3$ as a biosignature. However, \citet{segura10} suggest that if sufficient O$_3$ is present in a planet's atmosphere the majority of harmful UV radiation, even from a substantial flare, would be absorbed before reaching the surface, providing an effective shield for any life there. However, the impact of repeated flares on the atmosphere was not tested.

Whilst the case for ozone detection on Earth-like planets might not be so strong when differences in photochemistry are taken into account, the news might be better for a Venus analog planet. The Venusian clouds are made of sulphuric acid, which is produced photochemically at high altitudes. Different UV photon fluxes, in much the same way as they alter ozone production, may also inhibit (or enhance) sulphuric acid photolysis. This may either increase the likelihood of a Venus analog being relatively cloud-free, which makes detection of gas absorption features easier, or increase the production of sulphur aerosol, which may lead to an even flatter spectrum.

Finally, a persistent issue in transiting exoplanet studies is that the star must be very well characterised if the derived planet parameters are to be reliable, since all measurements are relative to the size and mass of the star. For many M stars, these properties are still not known to high precision. It will therefore be necessary to study an M dwarf primary in detail before attempting to characterize its companion.

\subsection{Abiotic false positives}
On the Earth, the observable abundance of O$_3$ arises directly from the large fraction of atmospheric O$_2$, which in turn arises from biological processes. However, observing a similar abundance of O$_3$ on an exoplanet would not necessarily imply biological activity, as there are several potential abiotic processes that could also generate substantial amounts of O$_2$. \citet{domagal14} model abiotic O$_2$, O$_3$ and CH$_4$ production from photolysis for planets orbiting a range of stars, and find that planets irradiated by an F2V star such as $\sigma$ Bootis would have O$_3$ column abundances only $\sim$7$\times$ smaller than present Earth's. Signatures of this size might be detectable in transit with \textit{JWST}. However, M dwarf host scenarios tested did not produce such large amounts of O$_2$ and O$_3$, with O$_3$ column abundances a factor of 1000 lower.  \citet{tian14} find that M dwarf hosts with high ($\sim$1000 super solar) FUV/UV flux ratios could also produce substantial amounts of O$_2$ and O$_3$ (O$_3$ abundances up to 0.05$\times$ present day Earth) without the requirement for biological processes. Since, as stated above, O$_3$ from Earth levels of biogenic oxygen could be substantially reduced around M dwarf hosts, distinguishing between the upper limit of abiotic O$_3$ and the lower bound of biologically related O$_3$ could potentially be very difficult.

\citet{luger15} postulate another abiotic production mechanism for substantial amounts of O$_2$. Earth-sized planets with a water ocean may lose those oceans through evaporation if they undergo a runaway greenhouse. Photodissociation of water vapour can then occur, and hydrogen is preferentially lost to space over the heavier oxygen. This can lead to a build-up of O$_2$ in the atmosphere, even up to several hundred bars. Whilst the authors do not discuss O$_3$ chemistry under these circumstances, it is conceivable that significant amounts could be produced. \citet{wordsworth14} point out that H$_2$O-rich atmospheres without significant amounts of non-condensing gas such as N$_2$ are particularly vulnerable to this process, as the non-condensing gas creates a cold trap for H$_2$ that aids retention. Other abiotic O$_2$ production mechanisms are also possible (e.g. \citealt{narita}). 

\subsection{Refraction and multiple scattering}
Currently, the \textit{NEMESIS} model does not account for refraction of transmitted light in the lower atmosphere of a transiting planet. This is relatively unimportant for hot Jupiters, which are high temperature and hydrogen rich, but the effect is enhanced for cooler planets. However, the size of the effect is also correlated with the size of the star, so it is less important for cool planets around small M dwarfs than similar planets around sun-like stars. \citet{betremieux14} find that there is very little difference between synthetic Earth spectra with and without refraction if the planet is assumed to be in orbit around a star between M5 and M9 spectral types. The differences are certainly far below the level of the noise on the synthetic spectra discussed in this work. By contrast, there are substantial differences for a true Earth-analog orbiting a sun-like star, further reinforcing the point that studying transiting Earth and Venus analogs around cooler stars is much more straightforward. 

For primary transit, \textit{NEMESIS} uses an extinction-only approximation for any scattering particles in the atmosphere. This is justified because of the extremely long path length in primary transit, which makes it likely that most photons encountering optically thick cloud will either be absorbed or scattered out of the beam. However, this does not take into account cloud particles with strongly forward-scattering phase functions. The sulphuric acid cloud on Venus is known to be strongly foward scattering, so it is possible that in the cloudy Venus calculation we are underestimating the amount of starlight that penetrates the atmosphere at pressures below the cloud top. \citet{dekok12} demonstrate this effect for Titan by comparing Monte Carlo photon-firing models that a) do not include multiple scattering and b) account for multiple scattering in strongly forward-scattering clouds. They find that the CH$_4$ and aerosol abundance is consistently underestimated for the non-multiple-scattering case, by a factor of around 0.92. However, we judge that the effects seen on the spectrum are likely to be small compared with the noise on the transmission spectra considered here. 

\subsection{Temporal Changes}
\label{temporal_changes}
So far, the transit spectroscopy technique has relied on it being possible to observe several transits of a planet and combine the data from each observation to increase the signal to noise ratio. This is a legitimate approach only if the stellar and atmospheric conditions are not expected to vary greatly between observations. In the case of an M dwarf, it is likely that star spots will be present, although with the relatively slow rotation period spot coverage should not change appreciably during a transit. However, given that several tens of observations are required for each planet, stellar activity may limit reliable combination of multiple transits. It may be possible to compensate for these effects if the star is monitored by an independent programme, as was done for the K star HD 189733 during the observational campaigns that resulted in the overview presented in \citet{pont13}. The effect of star spots decreases at longer wavelengths, so detection of the ozone feature would probably not be affected, but it would be much more difficult to place constraints on a Venus-like atmosphere where most obervable features are at wavelengths shorter than 5 $\upmu$m. 

When terrestrial planets are under consideration, it is also necessary to account for possible temporal variations in the planet's atmosphere, especially in cloud properties. Whilst the Venusian cloud layer is very stable, a Venus analog with reduced cloud opacity might not have such constant cloud cover. If the amount of cloud cover changes between observations, it might become impossible to combine spectra in any useful way as the size of absorption features would vary between observations - this can be seen in Figure~\ref{venusspec} comparing the top and bottom spectral plots.

The cloud cover on Earth also changes from day to day, although as the Earth cloud forms deeper in the atmosphere it has a smaller effect on the spectrum. However, certain events on Earth can cause dramatic changes in cloud and haze. For example, large volcanic eruptions such as Mount Pinatubo in 1992 are capable of increasing aerosol optical depth in the atmosphere on a global scale \citep{self97}, and this could have a significant effect on transmission spectra. Earth and Venus analogs around M5 stars would be in sufficiently close orbits to be within the tidal locking radius. Strong tidal forces may increase the rate of volcanic activity on these planets, possibly increasing the frequency of Pinatubo-size eruptions. 

In addition, the molecular features of interest could also vary in strength between observations. \citet{segura10} find that the O$_3$ column depth can vary by large amounts during a strong flare from an active M dwarf, especially when the effect of protons is accounted for. Attempting to average over repeated observations for a flaring M dwarf would therefore be problematic.

\begin{figure}
\centering
\includegraphics[width=0.5\textwidth]{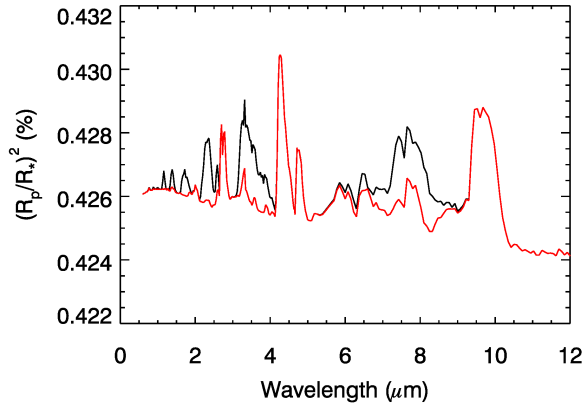}
\caption{The effect of a factor 200 increase in CH$_4$ abundance on the transmission spectrum for an Earth analog planet. The black line indicates a model with 200$\times$ CH$_4$, and the red line is the baseline Earth model. The 8$\upmu$m CH$_4$ band becomes almost as prominent as the 9 $\upmu$m O$_3$ feature.\label{ch4inc}}
\end{figure}

\section{Conclusions}
Whilst it is possible to identify strong absorption features, due to gases such as O$_3$, in \textit{JWST} transit measurements of M dwarf Earth analogs, there are many complicating factors. A reduced O$_3$ abundance compared with the Earth case, which seems to be likely considering the different UV flux for M stars compared with G stars, would make it harder to unambiguously identify the presence of ozone. The presence of high cloud on Venus analog planets means that gas absorption features could be almost completely obscured, making it very difficult to identify the atmospheric constituents and obfuscating attempts to characterise the planet. Less cloudy Venus-like planets, which may be more likely to occur around cooler stars, would be much more favourable for characterisation. 

More generally, characterising planets orbiting variable stars is difficult if multiple observations are needed to boost the signal. This is further complicated in the case of terrestrial planets, as their global atmospheric characteristics may also change on short timescales. Given the level of time investment required in observing this kind of target with \textit{JWST}, a very careful cost-benefit analysis would need to be completed for any such observation to be approved. Given the likelihood of signal disruption due to stellar activity or planet temporal variability, and the strong possibility that the planet will be cloudy, we recommend that the following criteria are fulfilled before such an observation is considered: 
\begin{itemize}
\item The Earth-like planet candidate is in a low-eccentricity orbit around a nearby ($<$ 10 pc) red dwarf with a spectral type of around M5
\item The host star is relatively quiet for its spectral type, with the caveat that stars with very low UV fluxes would be less likely to host planets with observable O$_3$
\item The host star can be periodically monitored for activity variations throughout the duration of \textit{JWST} observations
\item The target is the most favourable of its type, as it is likely that only 1--2 terrestrial planets can command this level of investment.
\item The mass and radius of both the planet and star have been measured to high precision (so the density of the planet is known).  
\end{itemize}

Although obtaining transit spectra of terrestrial planets is difficult, the first characterization of a cool, compact, terrestrial atmosphere would result in significant scientific return. If the criteria mentioned above can be met for an Earth-like planet candidate, we would advocate investing the required \textit{JWST} observing time. 

\section*{Acknowledgements}
JKB and PGJI acknowledge the support of the Science and Technology Facilities Council. LNF is supported by a Royal Society University Research Fellowship. We thank Victoria Meadows for pointers towards useful literature, and the anonymous reviewer whose constructive comments substantially improved the paper.

\bibliographystyle{mn2e}
\bibliography{bibliography}

\label{lastpage}
\end{document}